\documentclass[12pt,intlim,righttag]{article}
\usepackage{amsmath,amsthm}
\usepackage{amssymb}
\usepackage{amsfonts}

\newcommand{\td}{\tilde}

\newcommand{\lbl}{\label}

\newcommand\beq{\begin{equation}}
\newcommand\eeq{\end{equation}}

\newcommand{\bea}{\begin{eqnarray}}
\newcommand{\eea}{\end{eqnarray}}
\newcommand{\beaa}{\begin{eqnarray*}}
\newcommand{\eeaa}{\end{eqnarray*}}

\theoremstyle{Theorem}

\theoremstyle{corollary}

\theoremstyle{remark}

\theoremstyle{definition}

\def\a{\alpha}

\def\d{\delta}

\def\th{\theta}
\def\o{\omega}

\def\th{\theta}

\def\one{{\mathbf1}}

\usepackage{graphicx}
\newcommand{\comment}[1]{}

\begin{document}
\title{Bilateral Tariffs Under International Competition}

\author{T. Kutalia$^{1)}$ and R. Tevzadze$^{2)}$}

\date{~}
\maketitle
\begin{center}
$^{1)}$ Institute of Cybernetics,  5 Euli str., 0186, Tbilisi,
Georgia and
Georgian-American University, 8 Aleksidze Str., Tbilisi 0193, Georgia,
\newline(e-mail: tsotnekutalia@gau.edu.ge)
\\
$^{2)}$ Georgian-American University, 8 Aleksidze Str., Tbilisi 0193, Georgia,
Georgian Technical Univercity, 77 Kostava str., 0175,
Institute of Cybernetics,  5 Euli str., 0186, Tbilisi,
Georgia
\newline(e-mail: rtevzadze@gmail.com)
\end{center}

\begin{abstract}

This paper explores the gain maximization problem of two nations engaging in non-cooperative bilateral trade. Probabilistic model of an exchange of commodities under different price systems is considered. Volume of commodities exchanged determines the demand each nation has over the counter party's currency. However, each nation can manipulate this quantity by  imposing a tariff on imported commodities. As long as the gain from trade is determined by the balance between imported and exported commodities, such a scenario results in a two party game where  Nash equilibrium tariffs are determined  for various foreign currency demand functions and ultimately, the exchange rate based on optimal tariffs is obtained.

\end{abstract}

\bigskip

\noindent {\it 2010 Mathematics Subject Classification. 90A09, 60H30, 90C39}

\

\noindent {\it Keywords}: The Nash equilibrium,  tariff game, exchange rate

\section{Introduction}

Scientists have studied the trade gain maximization problem from different  perspectives. R. Gibbons \cite{gib} considered a game model in which total welfare of a country consists of an economic surplus enjoyed by consumers, profit earned by firms within a given country and the tariff revenue collected from the imports. Maximization of the total welfare from trade leads to optimal tariff countries involved in trade should impose.

 In \cite{LS}, a closed economy model is considered in which the country consists of a fixed number of households having preferences as a function of consumption and leisure. Within this model, consumption goods consist of intermediate goods that can be produced by units of labor. Under the closed economy model, quantities of each intermediate good and the tariff a given country imposes on imports are optimized.

Due to different circumstances of production, two nations can produce similar goods and services at different prices. They can both benefit by getting involved in international trade to import commodities, which under their own price system is of relatively low price than domestically produced commodities, which under the same price system is of relatively high price. The volume of commodities imported determines one nation's demand for another nation's currency. Balance of demands of two nations for foreign currency determines an exchange rate. J. T. Schwartz \cite{sch2} considered a model of gain maximization where the commodities produced and the prices for those commodities are static. In addition, gain from trade is determined to be the difference between the values of imported and exported commodities measured in national currency. Since importing those commodities which cost less under the national price system is regarded as a benefit for both nations, gain from competitive trade for a given nation is considered to be the difference between the advantage it took over the competitor and the advantage the competitor took over it, thus the difference between imports and exports measured at national currency. The Schwartz's model solves the tariff optimization problem for two nations which are said to be economically symmetric, meaning they have equal demands for each other's currency under a given exchange rate.

The novelty of the approach examined in this paper is to make the commodities and their prices random and solve the gain maximization problem under Nash's sense. In addition, non-cooperative bilateral trade model is generalized for asymmetric case and a more realistic problem where two nations have different demands for foreign currency is solved.

Greatest mutual benefit is achieved when nations cooperate and pursue a free trade policy. Here we assume the non-cooperative game, so they determine the optimal tariffs which results in greatest benefit for them under the Nash's sense.

\section{Equations for Nash-equilibrium tariffs and exchange rate}

Let us assume two nations exchange N different commodities for which the demand and prices are known. For the domestic and foreign nations, annual demand and corresponding prices measured in national currency are  $d_1,...,d_N, p_1,...,p_N$ and $d_1^*,...,d_N^*, p_1^*,...,p_N^*$ respectively. If we take $x$ as an exchange rate of a unit of foreign currency in terms of domestic currency units, then the domestic and foreign nations' demand for foreign currency are given by

\beq\lbl{demXD}
D(x):=\frac1{C_N}\sum_{k=1}^{N} \bar E\left(p_k^*d_k,\frac{p_k}{p_k^*}>x\right)
\eeq
and
\beq\lbl{demXF}
D^*(x):=\frac1{C_N^*}\sum_{k=1}^{N}\bar E\left(p_kd_k^*,\frac{p_k}{p_k^*}<x\right)
\eeq
respectively, where $C_N=\sum_{k=1}^{N}\bar E\left(p_k^*d_k\right)\;,C_N^*=\sum_{k=1}^{N}\bar E\left(p_kd_k^*\right)$ and $\bar E$
is the mathematical expectation under $\bar P$  on a probability space $(\bar \Omega,\bar F,\bar \Omega)$. If we introduce  the extended probability space
$(\Omega,F,P)$, where
$$\Omega = \bar \Omega \times\{1,...,N\},\;{P}(A,k)=\frac{1}{N}\bar P(A), A\in \bar F$$
and define random variables $p,p^*,d,d^*$ by
\beaa
p(\o,k)=p_k(\o),\;p^*(\o,k)=p_k^*(\o),\\
d(\o,k)=d_k(\o),\;d^*(\o,k)=d_k^*(\o),
\eeaa
then (\ref{demXD}),(\ref{demXF}) can be rewritten as probability distribution functions
\beq\lbl{disCondDF}
D(x)=E\left(p^*d,\frac{p}{p^*}>x\right),\;D^*(x)=E\left(pd^*,\frac{p}{p^*}<x\right).
\eeq
which indicate that the domestic nation will import the commodity if $\frac{p}{p^*}>x$ and the foreign nation will import if $\frac{p}{p^*}<x$.
Since $x$ is the value of a unit of foreign currency in terms of the domestic currency units, increasing the exchange rate makes foreign commodities more expensive for the domestic nation and the domestic commodities less expensive for the foreign nation. Therefore, $D$ is a decreasing function of $x$ and $D^*$ is an increasing function of $x$. These functions have the following properties

$$D(0)=1,\; D(\infty)=0, \;D^*(0)=0,\; D^*(\infty)=1.$$

\begin{figure}[!hbt]
  \centering
  \begin{minipage}[h]{0.4\textwidth}
    \includegraphics[width=\textwidth]{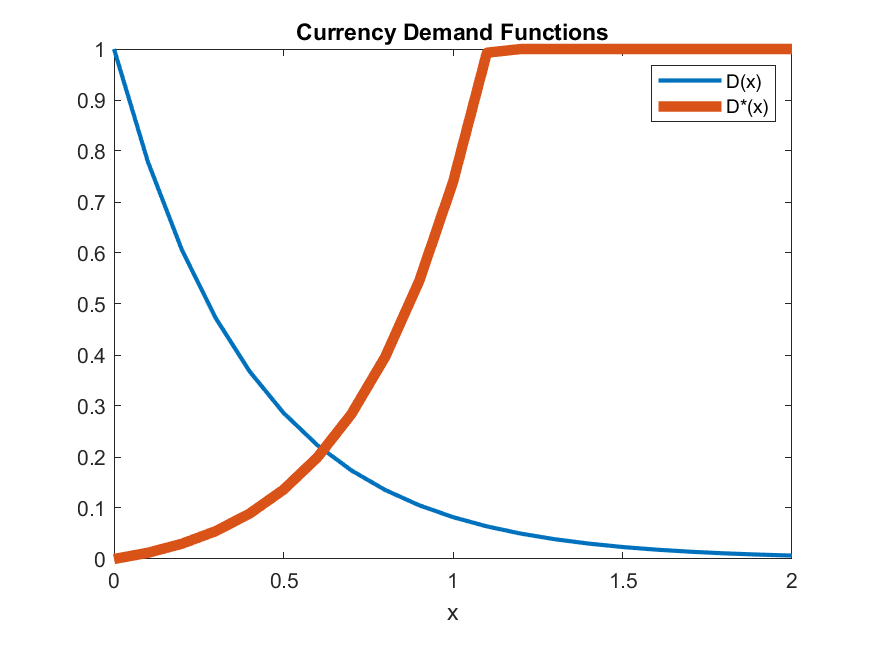}
    \caption{$D(x)$ and $D^*(x)$}
  \end{minipage}
  \hfill
  \begin{minipage}[h]{0.4\textwidth}
    \includegraphics[width=\textwidth]{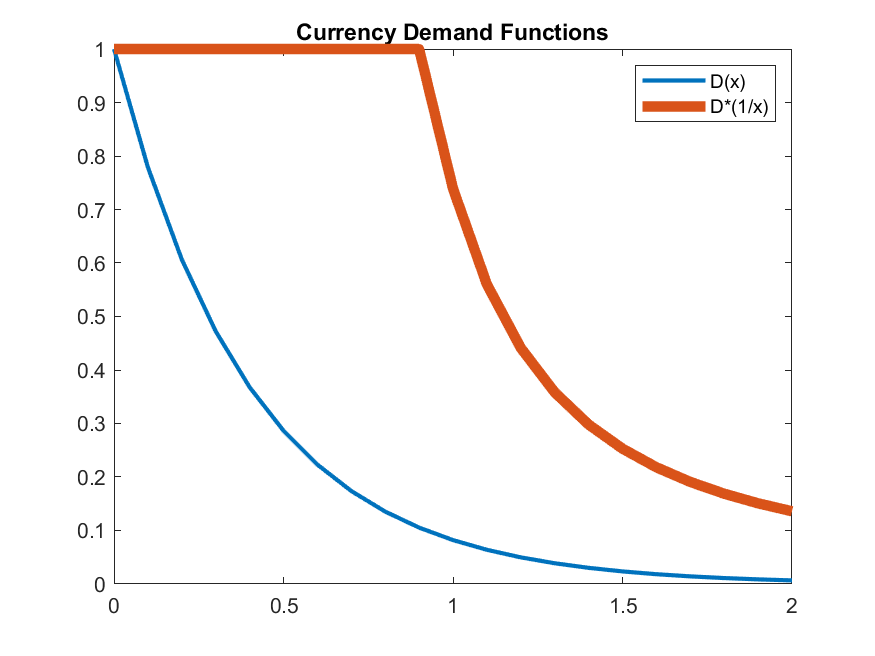}
    \caption{$D(x)$ and $D^*(1/x)$}
  \end{minipage}
\end{figure}

For an exchange rate $x$, solving the equation
\bea\lbl{balX}
xD(x)=D^*(x)
\eea
for $x$ yields the equilibrium rate $x=e$. This equation determines the equilibrium exchange rate when both nations practice an unrestricted free trade policy. Left side of the equation is the foreign currency demand of a domestic nation and the right side is the foreign currency demand of a foreign nation, both measured in domestic currency units.

Now suppose the domestic and foreign governments impose the following tariffs on imported commodities: $1-\th$ and $1-\th^*$. Then the domestic nation will import the commodity if $\frac{p\th}{p^*}>x$, and the foreign nation will import if $\frac{p^*\th^*}{p}>\frac{1}{x}$. Taking tariffs into account, the demand functions (\ref{disCondDF}) now become

\beq\lbl{demTDF}
D(\frac{x}{\th})=E\left(p^*d\one_{\{\th p>xp^*\}}\right),\;D^*(x\th^*)=E\left(pd^*\one_{\{\th^* p^*x>p\}}\right).\notag
\eeq
So the relation (\ref{balX}) is rewritten as
\beq\lbl{balT}
xD(\frac{x}{\th})=D^*(\th^*x),
\eeq
from which  it is clear that the equilibrium exchange rate $x=e$ now depends on $\th$ and $\th^*$. Equation (\ref {balT}) always has the solution $e=0,\;\frac{1}{e}=0$, or $\th=\th^*=0$, which do not carry any useful economic sense. Such conditions would restrict the involvement of both nations in trade. To rule out these possibilities, we claim $\frac1M\le e\le M$, for some large number $M$ and $\frac1M\leq\th\leq 1,\;\frac1M\leq\th^*\leq1$

Since the ultimate goal of both nations is to set the tariffs unilaterally  which will maximize their gain from trade, we have to find the Nash equilibrium point, the pair $(\hat \th, \hat \th^*)$. The gain functions of each nation are given by

\bea\lbl{gainD}
G(e,\th,\th^*)=E\left(pd,\frac{p^*}{p}<\frac\th e\right)-E\left(pd^*,\frac{p^*}{p}>\frac{1}{e\th^*}\right)\\
=E\left(\frac{p}{p^*}1_{(\frac{p}{p^*}>\frac{e}\th)}p^*d\right)-E\left(pd^*1_{(\frac{p}{p^*}>\frac{e}{\th})}\right)\notag\\
\notag =-\int_{e/\th}^\infty yD'(y)dy-D^*(\th^*e),
\eea
and
\bea\lbl{gainF}
G^*(e,\th,\th^*)=E(p^*d^*,\frac{p}{p^*}<\th^*e)-E(p^*d,\frac{p}{p^*}>\frac{e}{\th})\\
=\int_{\frac1{\th^*e}}^\infty\frac{1}{y}{D^*}'\left(\frac1y\right)dy-D\left(\frac{e}{\th}\right),\notag
\eea
respectively.
Since the equilibrium exchange rate is the function of tariffs, we have $e=e(\th,\th^*)$.
Our goal is to find the Nash equilibrium for the nations, i.e. such pair $(\hat\th,\hat\th^*)$ that relations
\bea
\max_\th G(e(\th,\hat\th^*),\th,\hat\th^*)=G(e(\hat\th,\hat\th^*),\hat\th,\hat\th^*),\notag\\
\max_{\th^*} G^*(e(\hat\th,\th^*),\th,\hat\th^*)=G^*(e(\hat\th,\hat\th^*),\hat\th,\hat\th^*)\notag
\eea
 hold.
The Nash pair is found from the system of equations
\bea\lbl{dgainD}
\frac{\partial}{\partial \th}G(e,\th,\th^*)=0,
\eea
\bea\lbl{dgainF}
\frac{\partial}{\partial \th^*}G^*(e,\th,\th^*)=0
\eea

Given the currency demand functions $D(x)$ and $D^*(x)$, solution to the system of equations (\ref{dgainD}),(\ref{dgainF}) leads to yet another system of equations (Appendix A)
\bea\lbl{sysMainOne}
D(\frac{e}{\th})=\th^*(1-\th){D^*}'(\th^*e),
\eea
\bea\lbl{sysMainTwo}
D(\frac{e}{\th})=\frac{e}{\th}(\th^*-1)D'(\frac{e}{\th})
\eea

{\bf Remark.} According to (\ref{balT}), $D(\frac{e}{\th})=\frac{D^*(\th^*e)}{e}$. Then (\ref{sysMainOne}) can be rewritten as
\bea\lbl{sysRedefOne}
D^*(\th^*e)=e\th^*(1-\th){D^*}'(\th^*e)
\eea
Denoting $\td e=\frac1e,\;\td D(x)=D^*(\frac1x)$,  (\ref{sysRedefOne}) now becomes
\bea\lbl{sysRedefTwo}
\td D(\frac{\td e}{\th^*})=\frac{\td e}{\th^*}(\th-1)\td D'(\frac{\td e}{\th^*}),\notag
\eea
which is similar to (\ref{sysMainTwo}).

At this point, if the demand functions for foreign currency of each nation are known, from (\ref{sysMainOne}) and (\ref{sysMainTwo}) the Nash equilibrium pair $(\hat \th,\hat \th^*)$ can be found.
Ultimately putting these values in (\ref{balT}) and solving for $x$ will result in the equilibrium triple $(\hat e, \hat \th, \hat \th^*)=(e(\hat \th, \hat \th^*), \hat \th, \hat \th^*)$.
Hence the triple satisfy
\bea\lbl{balTMain}
\hat eD(\frac{\hat e}{\hat\th})=D^*(\hat\th^*\hat e),
\\
\lbl{balDOne}
D(\frac{\hat e}{\hat \th})={\hat\th}^*(1-\hat\th){D^*}'(\hat\th^*\hat e),
\\
\lbl{balDTwo}
D(\frac{\hat e}{\hat\th})=\frac{\hat e}{\hat\th}(\hat\th^*-1)D'(\frac{\hat e}{\hat\th}).
\eea
Obviously, one should check whether the extremum points given by (\ref{sysMainOne}) and (\ref{sysMainTwo}) are really maximums.
Differentiating the derivatives of the gain functions once again and checking the signs for the equilibrium points serve this purpose.
So the following inequalities must hold
\bea\lbl{ddsys1}
\frac{\partial^2}{\partial \th^2}G(\hat e,\hat \th,\hat \th^*)<0,\notag
\eea
\bea\lbl{ddsys2}
\frac{\partial^2}{\partial \th^{*2}}G^*(\hat e,\hat \th,\hat \th^*)<0\notag
\eea
which means (Appendix A)
\bea\lbl{ine1}
\hat\th^{*2}(1-\hat\th) e_{\hat\th}{D^*}''(\hat\th^*\hat e)-\hat\th^*{D^*}'(\hat\th^*\hat e)-\frac{e_{\hat\th}\hat\th-\hat e}{{\hat\th}^2}D'(\frac{\hat e}{\hat\th})<0,
\eea
\bea\lbl{ine2}
\hat\th(\hat\th^*e_{\hat \th^*}+\hat e)D'(\frac{\hat e}{\hat\th})-(1-\hat\th^*)e_{\hat \th^*}\hat e D''(\frac{\hat e}{\hat\th})>0.
\eea
Hence we can formulate our main result:
{\it If pair $(\hat\th,\hat\th^*)\in(\frac1M,1)^2$ is a unique solution of (\ref{balTMain}), (\ref{balDOne}), (\ref{balDTwo}), (\ref{ine1}), (\ref{ine2}),
then it is the  Nash equilibrium of the game.}

\section{Examples of symmetric and asymmetric countries}

Demand functions differ from nation to nation. Specifically, two nations are said to be economically symmetric if
\bea\lbl{balSym}
D(x)=D^*\left(\frac{1}{x}\right),
\eea
which implies that their demand for each other's currency are equal under any given exchange rate. Economically, this means that on average they produce and exchange commodities of equal value. In case of symmetric nations, from (\ref{balT}) we can simply conclude that $e(\th^*,\th)=\frac1{e(\th,\th^*)}$. Thus $e(\th,\th)=1$, which makes perfect sense. Since two nations have equal demand for each other's currency, neither is able to employ dominant economic power over the counter party, so the Nash equilibrium will occur at equal tariffs and a unit exchange rate.
More rigorously, since $G^*(\frac1{e(\th,\th^*)},\th^*,\th)=G(e(\th,\th^*),\th,\th^*)$, from the result of Game Theory (\cite{MSZ},p.134) follows that $\hat\th=\hat\th^*$ for Nash point $(\hat\th,\hat\th^*)$.
This fact simplifies the computations above. Specifically, taking $\th=\th^*$ and $e=1$, (\ref{sysMainTwo}) becomes
\bea\lbl{sysSym}
\th D(\frac{1}{\th})=(\th-1)D'(\frac{1}{\th})
\eea
Given the function $D(x)$, the equilibrium pair $(\hat \th, \hat \th^*)$ is found.

\par
However, more realistic case is economically asymmetric nations having different demands for each other's currency. In this case, the equality (\ref{balSym}) no longer holds. So the nations will have different tariffs imposed on imported commodities.

In \cite{sch2}, the following demand functions for symmetric nations were considered: $D(x)=D^*(\frac1x)=(1+ x)^{-2}$. Since two nations are economically symmetric, we have $\hat \th= \hat \th^*, \hat e=1$. Use (\ref{sysSym}) to solve the equation for $\th$.
Given
$$D(x)=D^*(\frac{1}{x})=(1+x)^{-2}$$
the derivative of the function is
$$D'(x)=-\frac{2}{(1+x)^{3}}$$
putting it into (\ref{sysSym}) gives
$$\frac{\th}{(1+\frac{1}{\th})^2}=(\th-1)(-\frac{2}{(1+\frac{1}{\th})^3})$$
solving for $\th$ yields
$$\frac{-2\th}{1+\th}(\th-1)=1$$
$$\th=\frac{1}{3}=\th^*$$

The solution  $(\hat e,\hat\th,\hat\th^*)=(1,\frac{1}{3},\frac{1}3)$ agrees with Schwartz's results. So in case of symmetric nations with the foreign currency demand functions given by $D(x)$ and $D^*(\frac{1}{x})$, we obtained the Nash equilibrium point at equal tariffs to be imposed that maximize the gain for both nations from trade. However, neither is able to tax the competitor by a greater amount than itself being taxed by.

In addition, we consider two more examples.

\textbf{Symmetric case}.
\newline
Here we consider one more symmetric case. Suppose $D(x)=D^*(\frac1x)=(1-\a x)^+,\;\a<1$. Similarly applying (\ref{sysSym}) leads to the following solution.
Given
$$D(x)=D^*(\frac{1}{x})=(1-\alpha x)^+$$
the derivative is
$$D'(x)=-\alpha$$
putting it in (\ref{sysSym}) gives
$$\th(1-\frac{\alpha}{\th})=(\th-1)(-\alpha)$$
solving for $\th$
$$\th-\alpha=\alpha-\alpha\th$$
$$\th=\frac{2\alpha}{1+\alpha}=\th^*$$

So the Nash equilibrium point is $(\hat e,\hat\th,\hat\th^*)=(1,\frac{2\a}{1+\a},\frac{2\a}{1+\a})$. Similarly, given any value $\alpha<1$, which defines the shapes of the demand functions, the equilibrium point will occur at the same tariffs for both nations.

\textbf{Asymmetric case}.
\newline
Now we generalize the problem to a more common asymmetric case.Suppose $D(x)=\exp({-\delta x}),\;D^*(x)=(\alpha x\exp(\beta x))\wedge 1$. Then solving (\ref{balT}) yields the equilibrium exchange rate
$$e=\frac{-\th \ln(\alpha \th^*)}{\th \th^* \beta + \delta}$$
The Nash equilibrium condition (12),(13) gives $$\th=\frac{\delta(\th^*-1)\ln(\alpha\th^*) - \delta}{\th^*\beta}$$ and (Appendix B) $$\beta\th^*(\th^*-1)=(\th^*\beta-\delta(\th^*-1)\ln(\alpha\th^*)+\delta)(\th^*-(\th^*-1)\ln(\alpha\th^*)).$$ Specifically, if $\alpha=0.01, \beta=2, \delta=2.5$ the Nash equilibrium point is $(\hat e,\hat\th,\hat\th^*)=(0.81, 0.54, 0.73)$. The equilibrium exchange rate which is the solution of (\ref{balT}) is illustrated in the following figure:
\newline
\includegraphics[width=80mm,scale=0.5]{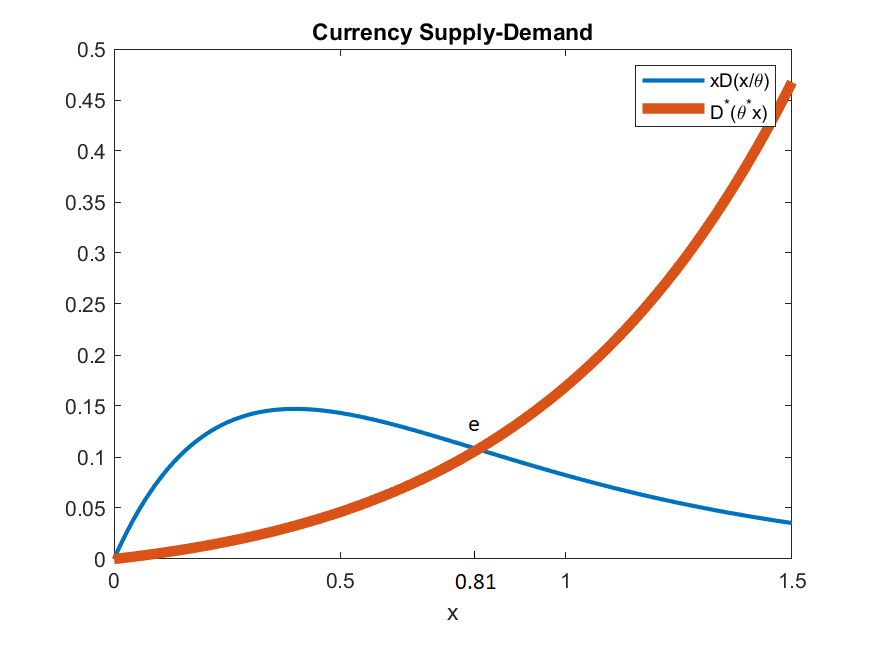}
\newline

Here the Nash equilibrium point is obtained for the case where $D(x)\neq D^*(1/x)$. Solution to the system of equations (\ref{sysMainOne}), (\ref{sysMainTwo}) leads to a domestic nation imposing greater tariff than the foreign nation. Initially, based only on the shapes of the functions $D(x)$ and $D^*(x)$, it is impossible to identify which nation is "economically stronger" and therefore will have a greater optimal tariff.

\section{Conclusion}

It can be concluded that the non-cooperative trade game results in a problem of optimizing tariffs on imported commodities. Optimal tariffs to be imposed are found at the Nash equilibrium point which are the solutions to the system of equations (\ref{balTMain}),(\ref{balDOne}),(\ref{balDTwo}). The shapes of demand functions determine the economic power one nation has over another. However, from the demand functions alone, it is impossible to predict which nation will have a greater optimal tariff to be imposed. It is assumed that the distributions of commodities exchanged and the prices for those commodities are known. Examples above are intended to illustrate the typical cases of economically symmetric and asymmetric nations involved in non-cooperative bilateral trade game. Obviously, in real world scenario, the demand functions are not predetermined. They are derived from (\ref{demXD}) and (\ref{demXF}). Finally, as an important note, it must be stressed that the vast improvement of the probabilistic model is that it does not restrict itself within the assumption of static prices and demand for commodities. Availability of imports and the tariffs affect the prices and demand for commodities. As long as they are made to be random, the model adequately responds to these changing conditions by considering different prices and demands on commodities.
\newline
\appendix {\bf Appendix A}\lbl{appA}
\newline

The following system of equations determine the tariffs each nation has to set in order to obtain the maximum gains from trade

\bea\lbl{dgainDA}
\frac{\partial}{\partial \th}G(e,\th,\th^*)=G_\th(e,\th,\th^*)+G_e(e,\th,\th^*)e_\th=0,
\eea
\bea\lbl{dgainFA}
\frac{\partial}{\partial \th}G^*(e,\th,\th^*)=G^*_{\th^*}(e,\th,\th^*)+G^*_e(e,\th,\th^*)e_\th=0,
\eea
where the index denotes the partial derivative of a given function with respect to a given variable.

If we define $g(e,\th)=\frac{e}{\th}$ and $g(e,\th^*)=\frac{1}{\th^*e}$, the individual components of (\ref{dgainDA}) and (\ref{dgainFA}) become

\bea\lbl{gainD_tA}
G_\th(e,\th,\th^*)=\frac{\partial}{\partial g}[-\int_{g}^\infty yD'(y)dy]\frac{\partial g}{\partial \th}-\frac{\partial}{\partial \th}D^*(\th^*e)
\\=gD'(g)g'(\th)\notag
\\=\frac{e}{\th}D'(\frac{e}{\th})(-\frac{e}{\th^2})=-\frac{e^2}{\th^3}D'(\frac{e}{\th}),\notag
\eea

\bea\lbl{gainD_eA}
G_e(e,\th,\th^*)=\frac{\partial}{\partial g}[-\int_{g}^\infty yD'(y)dy]\frac{\partial g}{\partial e}-\frac{\partial}{\partial e} D^*(\th^*e)
\\=gD'(g)g'(e)-\th^*{D^*}'(\th^*e)=\frac{e}{\th^2}D'(\frac{e}{\th})-\th^*{D^*}'(\th^*e),\notag
\eea

\bea\lbl{gainF_t1A}
G^*_{\th^*}(e,\th,\th^*)=\frac{\partial}{\partial g}[\int_{g}^\infty\frac{1}{y}{D^*}'\left(\frac1y\right)dy]\frac{\partial g}{\partial \th^*}-\frac{\partial}{\partial \th^*}D\left(\frac{e}{\th}\right)\
\\=-\th^*e{D^*}'(\th^*e)(-\frac{1}{\th^{*^2}e})=\frac{1}{\th^*}{D^*}'(\th^*e),\notag
\eea

\bea\lbl{gainF_eA}
G^*_e(e,\th,\th^*)=\frac{\partial}{\partial g}[\int_{g}^\infty\frac{1}{y}{D^*}'\left(\frac1y\right)dy]\frac{\partial g}{\partial e}-\frac{\partial}{\partial e}D\left(\frac{e}{\th}\right)
\\=-\th^*e{D^*}'(\th^*e)(-\frac{1}{\th^*e^2})-\frac{1}{\th}D'(\frac{e}{\th})=\frac{1}{e}{D^*}'(\th^*e)-\frac{1}{\th}D'(\frac{e}{\th})\notag
\eea

$e_\th$ and $e_{\th^*}$ can be found from (\ref{balT}) as follows. Let us define
\bea\lbl{funcF}
F(e,\th,\th^*)=eD(\frac{e}{\th})-D^*(\th^*e).
\eea
Differentiating (\ref{funcF}) with respect to $\th$ and $\th^*$ separately and equating them to zero yields the system of equations
\bea\lbl{funcF_t}
F_\th(e,\th,\th^*)+F_e(e,\th,\th^*)e_\th=0,\notag
\eea
\bea\lbl{funcF_t1}
F_{\th^*}(e,\th,\th^*)+F_e(e,\th,\th^*)e_\th^*=0,\notag
\eea
from which solving for $e_\th$ and $e_{\th^*}$ gives
\bea\lbl{et}
e_\th=-\frac{F_\th(e,\th,\th^*)}{F_e(e,\th,\th^*)}=\frac{\frac{e^2}{\th^2}D'(\frac{e}{\th})}{D(\frac{e}{\th})+\frac{e}{\th}D'(\frac{e}{\th})-\th^*{D^*}'(\th^*e)},\\
\lbl{et1}
e_{\th^*}=-\frac{F_{\th^*}(e,\th,\th^*)}{F_e(e,\th,\th^*)}=\frac{e{D^*}'(\th^*e)}{D(\frac{e}{\th})+\frac{e}{\th}D'(\frac{e}{\th})-\th^*{D^*}'(\th^*e)}
\eea
Putting these solutions into the system of equations (\ref{dgainDA})(\ref{dgainFA}) yields the following results. From (\ref{dgainDA}) we have
\bea
\frac{\partial}{\partial \th}G(e(\th,\th^*),\th,\th^*)\\
=-\frac{e^2}{\th^3}D'(\frac{e}{\th})
+[\frac{e}{\th^2}D'(\frac{e}{\th})-\th^*{D^*}'(\th^*e)]\left[\frac{\frac{e^2}{\th^2}D'(\frac{e}{\th})}{D(\frac{e}{\th})+\frac{e}{\th}D'(\frac{e}{\th})-\th^*{D^*}'(\th^*e)}\right]\notag
\\
=\frac{\frac{e^2}{\th^2}D'(\frac{e}{\th})}{D(\frac{e}{\th})+\frac{e}{\th}D'(\frac{e}{\th})-\th^*{D^*}'(\th^*e)}\notag\\
\times\left[{\frac{e}{\th^2}D'(\frac{e}{\th})-\th^*{D^*}'(\th^*e)}
-\frac{1}{\th}\left[{D(\frac{e}{\th})+\frac{e}{\th}D'(\frac{e}{\th})-\th^*{D^*}'(\th^*e)}\right]\right]\notag
\\
=\frac{e_\th}{\th^2}\left[eD'(\frac{e}{\th})-\th^2\th^*{D^*}'(\th^*e)-\th D(\frac{e}{\th})-eD'(\frac{e}{\th})+\th\th^*{D^*}'(\th^*e)\right]\notag
\\
=\frac{e_\th}{\th}[-D(\frac{e}{\th})+\th^*(1-\th){D^*}'(\th^*e)]=0\notag
\eea
and from (\ref{dgainF}) we have
\bea
\frac{\partial}{\partial \th^{*}}G^*(e(\th,\th^*),\th,\th^*)\\
=\frac{1}{\th^*}{D^*}'(\th^*e)+[\frac{1}{e}{D^*}'(\th^*e)-\frac{1}{\th}D'(\frac{e}{\th})][\frac{e{D^*}'(\th^*e)}{D(\frac{e}{\th})+\frac{e}{\th}D'(\frac{e}{\th})-\th^*{D^*}'(\th^*e)}]\notag
\\
=\frac{e{D^*}'(\th^*e)}{D(\frac{e}{\th})+\frac{e}{\th}D'(\frac{e}{\th})-\th^*{D^*}'(\th^*e)}\notag\\
\times\left[\frac{1}{e\th^*}[D(\frac{e}{\th})+\frac{e}{\th}D'(\frac{e}{\th})-\th^*{D^*}'(\th^*e)]
+[\frac{1}{e}{D^*}'(\th^*e)-\frac{1}{\th}D'(\frac{e}{\th})]\right]\notag
\\
=
\frac{e_{\th^*}}{e}\left[\frac{1}{\th^*}D(\frac{e}{\th})+\frac{e}{\th^*\th}D'(\frac{e}{\th})-{D^*}'(\th^*e)
+{D^*}'(\th^*e)-\frac{e}{\th}D'(\frac{e}{\th})\right]\notag\\
=\frac{e_{\th^*}}{\th^*e}[D(\frac{e}{\th})-\frac{e}{\th}(\th^*-1)D'(\frac{e}{\th})]=0\notag
\eea
Hence
\bea
D(\frac{e}{\th})=\th^*(1-\th){D^*}'(\th^*e),\notag\\
D(\frac{e}{\th})=\frac{e}{\th}(\th^*-1)D'(\frac{e}{\th}).\notag
\eea
For the second derivatives we have
\bea
\frac{\partial^2}{\partial \th^2}G(e(\th,\th^*),\th,\th^*)\notag\\
=\frac{\th e_{\th\th}-e_\th}{\th^2}[\th^*(1-\th){D^*}'(\th^*e)-D(\frac{e}{\th})]\notag\\
\lbl{summ1}+\frac{e_\th}{\th}\left[{\th^*}^2(1-\th)e_\th{D^*}''(\th^*e)-\th^*{D^*}'(\th^*e)-\frac{e_{\th}\th-e}{\th^2}D'(\frac{e}\th)\right],\\
\frac{\partial^2}{\partial \th^{*2}}G^*(e(\th,\th^*),\th,\th^*)\notag\\
=\frac{e\th^*e_{\th^*\th^*}-\th^*e_{\th^*}^2-e e_{\th^*}}{(e\th^*)^2}[D(\frac{e}{\th})-\frac{e}{\th}(\th^*-1)D'(\frac{e}{\th})]\notag\\
+\frac{e_{\th^*}}{e\th^*}\left[\frac{e_{\th^*}}{\th}D'(\frac{e}\th)-\frac{(1-\th^*)e_{\th^*}-e}{\th}D'(\frac{e}\th)-\frac{(1-\th^*)e_{\th^*}e}{\th^2}D''(\frac{e}\th)\right]\notag\\
=\frac{e\th^*e_{\th^*\th^*}-\th^*e_{\th^*}^2-e e_{\th^*}}{(e\th^*)^2}[D(\frac{e}{\th})-\frac{e}{\th}(\th^*-1)D'(\frac{e}{\th})]\notag\\
\lbl{summ2}
+\frac{e_{\th^*}}{e\th^*}\left[\frac{\th^*e_{\th^*}+e}{\th}D'(\frac{e}\th)-\frac{(1-\th^*)e_{\th^*}e}{\th^2}D''(\frac{e}\th)\right].
\eea
Assuming that the system (\ref{sysMainOne}), (\ref{sysMainTwo}) has a unique solution and $xD(x)\to 0,\;as\; x\to\infty$, we get
\beaa
  xD(\frac x\th)-D^*(\th^*x) >0, \;\mbox{if}\; x<e, \\xD(\frac x\th)-D^*(\th^*x) =0,\; \mbox{if}\; x=e, \\ xD(\frac x\th)-D^*(\th^*x)<0,\; \mbox{if}\; x>e.
\eeaa
Then
$$\frac d{dx}\bigg|_{x=e}  (xD(\frac x\th)-D^*(\th^*x))=D(\frac{e}{\th})+\frac{e}{\th}D'(\frac{e}{\th})-\th^*{D^*}'(\th^*e)<0$$
and from (\ref{et}),(\ref{et1}) follows that $e_\th>0,\;e_{\th^*}<0$. Since first summands of (\ref{summ1}),(\ref{summ2}) are zeros, the conditions $\frac{\partial^2}{\partial \th^2}G(e(\th,\th^*),\th,\th^*)<0, \;
\frac{\partial^2}{\partial \th^{*2}}G^*(e(\th,\th^*),\th,\th^*)<0$  provide (\ref{ine1}),({\ref{ine2}).
\newline
\appendix {\bf Appendix B}\lbl{appB}
\newline

Differentiating the given demand functions
\bea\lbl{d*}D(x)=\exp(-\delta x),\; D^*(x)=\alpha x\exp(\beta x)
\eea
gives
\bea\lbl{der}
D'(x)=-\delta\exp(-\delta x), \;{D^*}'(x)=(\alpha\beta x+\alpha)\exp(\beta x),\\
D''(x)=\delta^2\exp(\delta x),\; {D^*}''(x)=(\alpha\beta^2 x+2\alpha\beta)\exp(\beta x)
\eea
The equilibrium exchange rate is found from (\ref{balT}) as follows
\bea\lbl{exr}
e\exp(-\delta\frac{e}{\th})=\alpha\th^*e\exp(\beta\th^*e),\\
-\delta\frac{e}{\th}=\ln(\alpha\th^*)+\beta\th^*e,\notag\\
e(\th\th^*\beta+\delta)=-\th\ln(\alpha\th^*),\notag\\
\lbl{eSol}
e=\frac{-\th\ln(\alpha\th^*)}{\th\th^*\beta + \delta}.
\eea
From (\ref{sysMainTwo}) we find $\th$

$$\exp(-\delta\frac{e}{\th})=\frac{e}{\th}(\th^*-1)(-\delta\exp(-\delta\frac{e}{\th})),$$
$$1=\frac{e}{\th}(1-\th^*)\delta,$$
$$1=\frac{(\th^*-1)\delta\ln(\alpha\th^*)}{\th\th^*\beta+\delta},$$
\bea\lbl{thSol}
\th=\frac{(\th^*-1)\delta\ln(\alpha\th^*)-\delta}{\th^*\beta}.
\eea
Putting (\ref{eSol}),(\ref{thSol}) into (\ref{sysMainOne}) leads to the solution of $\th^*$. Specifically, redefining (\ref{sysMainOne}) in terms of (\ref{d*}) gives
$$\exp(-\delta\frac{e}{\th})=\th^*(1-\th)\alpha\exp(\beta\th^*e)(\beta\th^*e+1),$$
putting (\ref{eSol}) into this equation results in the following expression
$$\exp(\delta\frac{\ln(\alpha\th^*)}{\th\th^*\beta+\delta})=\th^*(1-\th)\alpha\exp(\beta\th^*[\frac{-\th\ln(\alpha\th^*)}{\th\th^*\beta+\delta}])(\beta\th^*[\frac{-\th\ln(\alpha\th^*)}{\th\th^*\beta+\delta}]+1),$$
replacing $\th$ with its definition from (\ref{thSol})
\begin{align*}
\exp(\frac{\delta\ln(\alpha\th^*)}{(\th^*-1)\delta\ln(\alpha\th^*)-\delta+\delta})
\\
=\th^*\frac{\th^*\beta-(\th^*-1)\delta\ln(\alpha\th^*)+\delta}{\th^*\beta}\alpha\exp(\frac{1-(\th^*-1)\ln(\alpha\th^*)}{\th^*-1})(\frac{1-(\th^*-1)\ln(\alpha\th^*)}{\th^*-1}),
\end{align*}
eliminating and rearranging some terms gives a simplified equation
\begin{align*}
\exp(\frac{1}{\th^*-1})\\
=\frac{\th^*\beta-(\th^*-1)\delta\ln(\alpha\th^*)+\delta}{\beta}\alpha\exp(\frac{1-(\th^*-1)\ln(\alpha\th^*)}{\th^*-1})\frac{\th^*-(\th^*-1)\ln(\alpha\th^*)}{\th^*-1},\\
\end{align*}
combining the exponents gives
\begin{align*}
\exp(\frac{1}{\th^*-1}-\frac{1-(\th^*-1)\ln(\alpha\th^*)}{\th^*-1})\\
=\frac{\th^*\beta-(\th^*-1)\delta\ln(\alpha\th^*)+\delta}{\beta}\alpha\frac{\th^*-(\th^*-1)\ln(\alpha\th^*)}{\th^*-1},\\
\end{align*}
simplifying the power of the exponent yields
\begin{align*}
\alpha\th^*=\frac{\th^*\beta-(\th^*-1)\delta\ln(\alpha\th^*)+\delta}{\beta}\alpha\frac{\th^*-(\th^*-1)\ln(\alpha\th^*)}{\th^*-1},
\end{align*}
finally, we obtain the equation involving only $\th^*$ to solve for
\bea\lbl{th1Sol}
\beta \th^*(\th^*-1)=(\th^*\beta-(\th^*-1)\delta\ln(\alpha\th^*)+\delta)(\th^*-(\th^*-1)\ln(\alpha\th^*)).
\eea
This equation cannot be explicitly solved for $\th^*$ but it can be computed approximately. Putting $\alpha=0.01,\beta=2,\delta=2.5$ into (\ref{th1Sol}) gives $\th^*=0.73$, putting this value into (\ref{thSol}) gives $\th=0.54$, and ultimately the equilibrium exchange rate is obtained by putting these values in (\ref{eSol}) which gives $e=0.81$. So the equilibrium tripe is $(\hat e,\hat \th, \hat \th^*)=(0.81, 0.54, 0.73)$.

The derivatives of the exchange rate function with respect to $\th^*$ and $\th$ are
\beaa
e_{\th^*}=\frac{\beta\th^2\th^*\ln(\alpha\th^*)-\th(\th\th^*\beta+\delta)}{\th^*(\th\th^*\beta+\delta)^2}=-0.49,\\
e_{\th}=\frac{-\delta \ln(\alpha\th^*)}{(\th\th^*\beta+\delta)^2}=1.13.
\eeaa
Using (\ref{der})-(\ref{exr}), inequalities (\ref{ine1}),(\ref{ine2}) take the form
\beaa
\th^*(1-\th)e_\th(\th^*e\beta^2+2\beta)-\th^*e-1+\frac{\delta}{\th^2}(e_\th\th-e)<0,\\
\th(\th^*e_{\th^*}+e)+\d (1-\th^*)ee_{\th^*}<0.
\eeaa
For $(\hat e,\hat e_{\th},\hat e_{\th^*},\hat \th, \hat \th^*)=(0.81,1.13,-0.49,0.54, 0.73)$, these inequalities can be verified.

\end{document}